# High-Performance and Distributed Computing in a Probabilistic Finite Element Comparison Study of the Human Lower Leg Model with Total Knee Replacement

*Corneliu T.C. Arsene*
School of Engineering Sciences
University of Southampton, SO17BJ
Southampton, United Kingdom
CorneliuArsene@gmail.com

*Abstract*—Reliability theory is used to assess the sensitivity of a passive flexion and active flexion of the human lower leg Finite Element (FE) models with Total Knee Replacement (TKR) to the variability in the input parameters of the respective FE models. The sensitivity of the active flexion simulating the stair ascent of the human lower leg FE model with TKR was presented before in [1,2] whereas now in this paper a comparison is made with the passive flexion of the human lower leg FE model with TKR. First, with the Monte Carlo Simulation Technique (MCST), a number of randomly generated input data of the FE model(s) are obtained based on the normal standard deviations of the respective input parameters. Then a series of FE simulations are done and the output kinematics and peak contact pressures are obtained for the respective FE models (passive flexion and/or active flexion models). Seven output performance measures are reported for the passive flexion model and one more parameter was reported for the active flexion FE model (patello-femoral peak contact pressure) in [1]. A sensitivity study will be performed based on the Response Surface Method (RSM) to identify the key parameters that influence the kinematics and peak contact pressures of the passive flexion FE model. Another two MCST and RSM-based probabilistic FE analyses will be performed based on a reduced list of 19 key input parameters. In total 4 probabilistic FE analyses will be performed: 2 probabilistic FE analyses (MCST and RSM) based on an extended set of 78 input variables and another 2 probabilistic FE analyses (MCST and RSM) based on a reduced set of 19 input variables. Due to the likely computation cost in order to make hundreds of FE simulations with MCST, a high-performance and distributed computing system will be used for the passive flexion FE model the same as it was used for the active flexion FE model in [1].

I. INTRODUCTION

This Finite Element (FE) studies have been used for orthopaedic problems in computational biomechanics for more than 20 years [3-5]. The majority of the studies have focussed on the implanted proximal femur but subsequently, there has been considerable interest in modelling of Total Knee Replacement (TKR). Modelling the behaviour of TKR is challenging, as the stresses generated within the prosthesis and the supporting bone are a function of the kinematics, and the kinematics in turn are a function of the implant design, relative position of the components (with respect to each other and with respect to the bones) and the balance of the soft tissues. Clinical studies [6,7] have shown that the kinematics of TKR are highly variable and a potential cause of this is variability in implant positioning.

For computer studies, in order to assess the likely performance of an implant, all of the input variability needs to be incorporated into a suitable computational model [8-10]. Laz et al. [11,12] have pioneered the application of probabilistic FE analyses to assess the performance of TKRs. They used the Advance Mean Value (AMV) method together with a 1000 run Monte Carlo Simulation Technique (MCST) for assessing the impact of experimental variability in the Stanmore knee wear simulator on predicted TKR mechanics by determining the envelope of performance of joint kinematics and contact parameters. Combination of the FE solvers with probabilistic methods (i.e. probabilistic FE analyses) have been applied to reliability of other orthopaedic components such as hip replacements [13-16] and for a longer time in the assessment of structural reliability [17-19]. This combination results in a more rigorous validation of the computation model and a more realistic comparison of the predicted results with results obtained by other means such as experimental results [11,12].

Providing a more complicated setup than the isolated TKR mode, the passive flexion FE lower limb model with TKR is studied in this paper. A brief comparison is made also with previous studies [1,2] involving an active flexion FE human lower limb model with TKR. The probabilistic FE studies presented in this paper will benefit especially from a powerful high-performance and distributed computing platform.

## II. ACTIVE/PSSIVE FLEXION FINITE ELEMENT MODELS WITH TOTAL KNEE REPLACEMENT

The passive flexion lower limb model consists of three main modelling parts bones, TKR components and soft tissues (Fig.1a). Preliminary loading and boundary conditions are applied through the hip and ankle joints, as well as a quadriceps force for the passive flexion, ensuring that the knee is kinematically unconstrained. The model includes the pelvis, femur, tibia and fibula. These bones are modelled using shell elements and defined as a rigid body. In the stochastic study, the positions of the center of hip and the centre of ankle are modified for three degrees of freedom (x, y and z).

The mean values and the standard deviations of the normal distributions of the 77 input variables of the passive flexion FE model from which the sampling process is done are shown in [1] for the active flexion. To the respective list is added in the passive flexion model the quadriceps load variable which forms 78 input variables in total. The variability of quadriceps load used in the probabilistic FE analyses in this paper is 10%.

Kinematic joints are modelled at the hip and the ankle to replicate the soft tissue constraints that are in place in vivo. Each joint is fixed in the translational degrees of freedom but free in the rotational degrees of freedom. The stiffness and the friction of the two kinematic joints will be varied in the stochastic study. Figure 1 shows the implanted TKR which is a PFC Sigma (DePuy) posterior cruciate sacrificing device. The position of the femoral, tibial and patellar components that are modelled using rigid shell elements, as described above, will be varied for all 6 degrees of freedom (3 translations and 3 rotations) in the probabilistic FE studies.

For each of the three components, the rotations will be applied to a node called centre of gravity which will be attached to the respective components. A penalty-based method was employed to define contact [20]. Contact is modelled between the femoral and tibial components (tibio-femoral contact) and the patellar and femoral components (patello-femoral contact). Friction is modelled between the contacting components and is given a value of 0.04 [20, 21]. The friction parameter will be varied in the probabilistic FE study. The final part included in the FE models is the soft tissues. These are currently modelled as simple non-linear tension-only bars, which can be seen in Fig.1. The soft tissues are created by modelling a nonlinear tension-only bar element between 2 nodes, one on each rigid body. A single bar is used to represent the Medial and Lateral Collateral Ligaments (MCL and LCL) and medial and lateral patellar retinaculum, whilst three bars in parallel model the quadriceps tendon and the patellar ligament. The positions of the insertion nodes for the MCL and LCL will be varied in the probabilistic FE study for 3 degrees of freedom (x, y, z). The material used for ligaments has zero resistance in compression and behaves in a nonlinear elastic-plastic fashion in tension. Additionally, these elements are unable to transmit bending and torsion moments. The MCL, LCL and patellar retinaculum are modelled with an initial pretension (pre-strain) in order to replicate their properties in vivo. The material properties (stiffness, pre-strain) of the collateral ligaments will be varied in the probabilistic FE study. The stiffness and two initial rotations of the quadriceps around the medial lateral axis and around the anterior posterior axis will be varied as well.

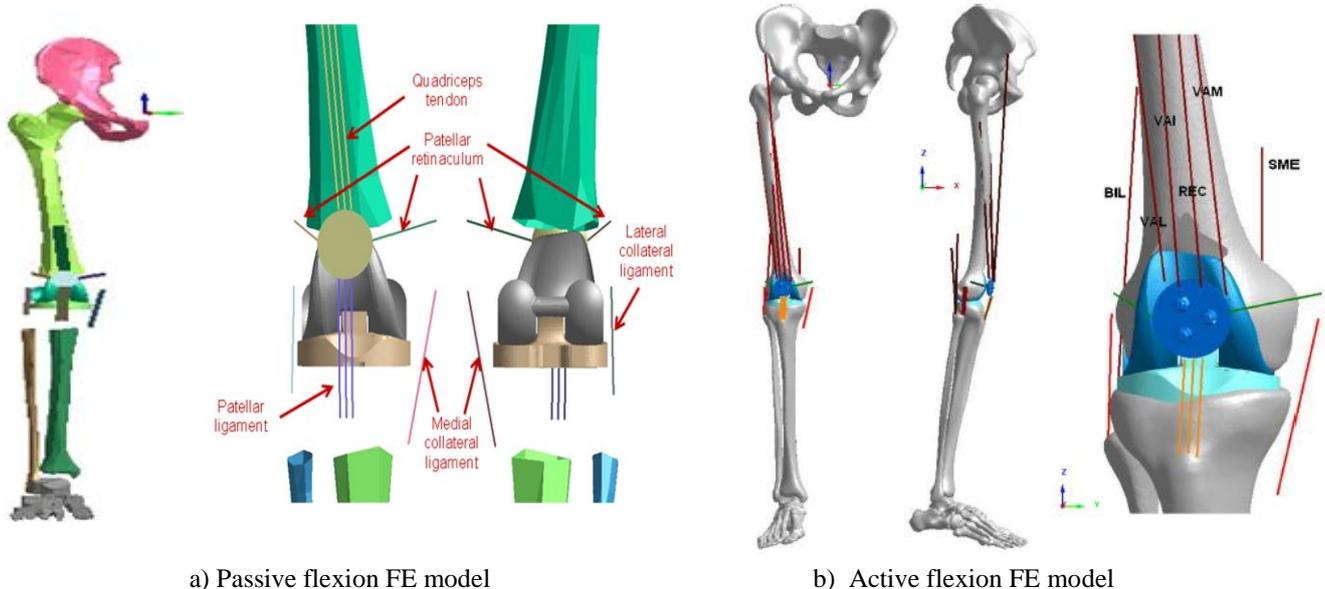

a) Passive flexion FE model        b) Active flexion FE model

**Fig. 1.** a) Passive flexion FE model of the human lower limb includes bones with TKR component, the soft tissues modelled as tension bars, patellar ligament, quadriceps tendon, MCL and LCL ligaments, the patellar retinaculum; b) Active flexion FE model of the human lower limb where forces are applied (i.e. FE model and probabilistic FE analyses presented in detail in [1]).

Also the friction of the patellar ligament and the quadriceps tendon will be varied. Forces or displacements can be applied to the model to generate motion. In this paper, a passive flexion cycle is modelled while in [1] the active flexion was presented in detailed. Passive flexion is a clinically relevant movement, as during TKR surgery the surgeon will passively flex the limb in order to check the implanted components. The flexion cycle lasts for 1000 ms. Passive flexion is achieved by applying two functions to the lower limb: firstly, a small quadriceps load of 50N is applied that always acts along the mechanical axis of the femur, regardless of the orientation of the femur in the global coordinate system. The FE model has a settling phase of 100 ms during which the quadriceps load is not applied. The quadriceps load will be varied in the probabilistic FE study. Secondly, motion is achieved by applying a 100N force to the ankle that acts in the direction of the hip. This results in a flexion of the knee joint of approximately 135°. The force is applied after 500 ms.

### III. PROBABILISTIC FINITE ELEMENT STUDIES OF THE ACTIVE/PASSIVE FLEXION FINITE ELEMENT MODELS WITH TOTAL KNEE REPLACEMENT

There will be two reliability techniques used in the probabilistic FE study of the passive flexion FE lower limb model. The effect of the full 78 input variables, from which 77 input variables are presented in [1], have on the kinematics and contact pressure will be investigated. Two probabilistic methods will be used: the MCST and the Response Surface Method (RSM). Following a sensitivity study, a reduced set of input variables will be derived, which will represent the key parameters for the passive flexion lower limb model. The reduced set will be investigated in a second probabilistic FE study. The aim is first to find out how well the reduced set can approximate the numerical results obtained with the full set of input variables as well as whether there are any changes in the envelopes of performance for the results obtained with the MCST and the same reduced set.

The first probabilistic method used is MCST. For the MCST method, each input variable assumes a normal probability distribution from which a number of random samples are generated. A set of samples are used in the FE simulation to obtain output values. The sampling process and the FE simulations are repeated until a wide sample of output results are obtained.

The second probabilistic method used in this paper is the RSM [22]. The RSM fits a mathematical function of the input variables (called the Response Surface Equation, RSE) to approximate the output parameter, across the full range of the sample space. Typically, this will be a low-order polynomial, and regression techniques are used to select the term coefficients. The method comprises three steps: first, a response vector $y$ (i.e. AP translation, IE rotation, Peak Contact Pressure) is obtained from a probabilistic FE simulation based on a sample of input conditions, denoted by vector X (the number of point will be smaller than that needed for the MCST method; in this case no more than 100 trials).

Trials could be random, but a better result is achieved by distributing the trials regularly across the sample space. Secondly, with the least squares method we can represent the results as a system of equations and arrange thus:

$$b = \left(X^T X\right)^{-1} X^T y \qquad (1)$$

where $b$ denotes the coefficients of the RSE.

Third, the RSE ($b$) together with a large number of Gaussian distributed samples (e.g. 1000) for each input variable forming matrix $X_1$ will generate the response vector of interest $y_1$:

$$y_1 = b X_1 \qquad (2)$$

Similar statistical measures can be applied to the response vector $y_1$ (e.g. mean, range, standard deviation, etc). This method works best when the true output is well represented by the mathematical function, for example relatively linear, smooth and monotonic models can easily be fitted [24,25]. It is expected that the passive flexion lower limb model will be relatively linear when a reduced set of input variables will be used.

Finally a sensitivity study is performed as described in [1] where a sensitivity matrix is calculated for each output measure, which sensitivity matrix has the number of rows equal to the number of input variables and the number of columns equal to the number of samples used to calculate Eq. 1. The matrix $X_1$ of each output kinematics and peak contact pressure variable is divided by the standard deviations σ of the input variables:

$$A = b \frac{X_1}{\sigma} \qquad (3)$$

where A is the sensitivity matrix with the number of rows equal to the number of input variables and the number of columns equal to the number of Monte Carlo samples (e.g. 50, 100) used in equation (2).

The sensitivity of an output variable $s_{ji}$ (e.g. tibial anterior-posterior translation) with respect to an input variable $i$ is calculated as the mean of the absolute values of the input variable row $i$ of the sensitivity matrix $A_j$ corresponding to the respective output variable $j$.

$$s_{ji} = mean(abs(A_j(i,:))) \qquad (4)$$

where $s_{ji}$ is the sensitivity of j output variable with respect to the input variable i, j is an index with values from 1 to 7 for each output variable, i is an index with values from 1 to 78 for the probabilistic FE analysis with RSM based on extended set of input variables or with values from 1 to 19 for the probabilistic FE analysis with RSM based on the reduced set of input variables, Aj is the j-th matrix which corresponds to the j output variable.

## IV. HIGH-PERFORMANCE AND DISTRIBUTED COMPUTING FOR COMPUTATIONAL BIOMECHANICS

The optimization/distributed computing software PamOpt (ESI, Paris) will be used together with the FE software PamCrash [26,27] to implement the MCST and the RSM. PamOpt is able to call PamCrash processes in parallel on several Personal Computers (PCs) assuming that PamCrash/PamOpt software was installed before on the respective PCs. Fig.2 (courtesy of ESI, Paris) shows how PamOpt makes simultaneous calls on remote platforms. This way it is possible to drastically reduce [1] the computational times required to obtain the envelope of performance for the kinematics and the contact pressure of the active/passive flexion FE lower limb models. This forms a high-performance and distributed computing platform for computational biomechanics. In order to implement the distributed computing framework formed by PamOpt/PamCrash software the Cygwin software [23] had to be installed and configured so that to be able to run the native Linux application PamOpt on Windows. There were used 2 PCs with the following specifications: a) Intel(R) Core(TM)2 CPU 6600 @ 2.40 GHz 2.40 GHz and b) Intel(R) Pentium(R) 4 CPU 3.20 GHz 3.19 GHz. The MCST points were distributed on the two different PCs and the computational time was reduced. While one passive flexion FE simulation, took about 50 minutes to complete, 800 points were obtained in 11 days by using in parallel the 2 PCs described above. Obviously further reduction in computational time can be achieved by adding more PCs in parallel which is not regarded as a problem at the present cost of a PC. For example, adding one more PC of type (a) in parallel would reduce the computational time to approximately 6 days and with more PCs added in parallel (i.e. 3 or 4) the computational time would be drastically reduced to even few hours. A further reduction in time would be achieved by modifying the characteristics of the passive FE model in order to run far less than 50 minutes as for example in under 10 minutes.

## V. NUMERICAL RESULTS AND CONCLUSIONS

The kinematics of the tibio-femoral contact joint and the patello-femoral contact joint were reported in Grood and Suntay joint coordinate system.

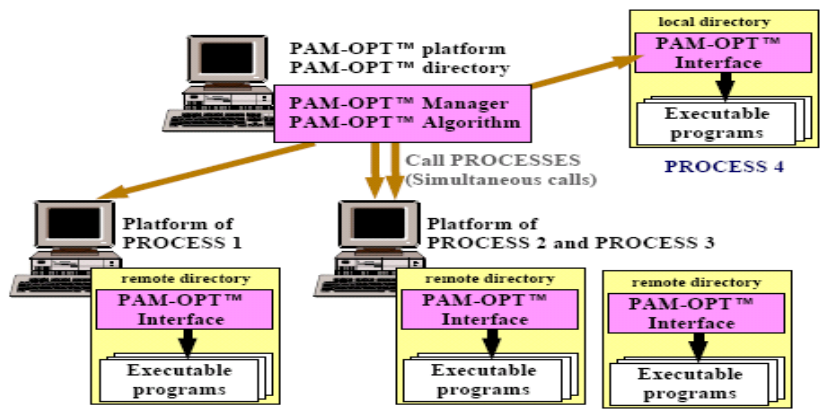

**Fig.2.** PamOpt makes simultaneous calls on remote platforms of the FE software PamCrash (courtesy of ESI, Paris).

The following 7 output performance metrics were considered for the passive flexion: tibio-femoral flexion angle, tibio-femoral peak contact pressure, tibial anterior-posterior translation, tibial internal-external rotation, patello-femoral flexion angle, patellar medial-lateral displacement, patellar tilt. During the MCST, it was looked to the 5% and 95% percentile (i.e. the values which are below, respectively above the respective percentiles were taken out), and the mean value of the peak value of the tibio-femoral flexion angle. It was observed that after 400 FE simulations the mean and the percentile values became constant (Fig. 3).

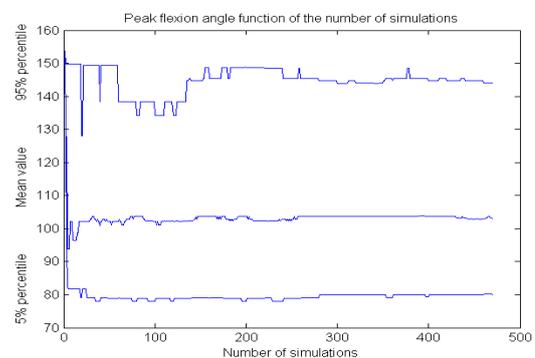

**Fig.3** Peak flexion angle function of the number of simulations for the probabilistic FE analysis based on MCST with the full set of 78 input variables.

However, there are used 800 FE simulation points after which the probabilistic FE analysis based on the MCST was stopped. Furthermore the RSM with 800 FE simulation points obtained a second set of envelopes which compared very well (i.e not shown in the paper) with the performance envelopes obtained with MCST with 800 FE simulations. With the RSM were also calculated the sensitivity coefficients. The most important 19 key parameters which influence the output kinematics and the peak contact pressure of the passive flexion model were calculated as described in [1] for the active model and in the previous section. The key parameters were calculated by summing their ranking positions from the seven lists of 78 input variables. After the 19$^{th}$ key parameter a significant dropped was noticed in the total scores of the key parameters.

The 19 key input parameters are friction of the pelvis-hip joint on the r-rotational direction (anterior-posterior: coordinates 0.71 (x) and 34.56 (y)), anterior-posterior position of the patellar component, quadriceps load, quadriceps initial rotation around the medial lateral direction, quadriceps initial rotation around the anterior-posterior direction, varus-valgus position of the femoral component, varus-valgus position of the tibial component, inferior-superior position of the femoral component, stiffness of the LCL ligament, friction of the leg ankle on the r-rotational direction (anterior-posterior: coordinates 0.71 (x) and 34.56 (y)), inferior-superior position of the tibial component, tilt of the tibial component, friction of the pelvis-hip joint on the r-rotational direction (anterior-posterior: coordinates 0.289 (x) and y (2.41)), internal-external position of the patellar component, x coordinate of the insertion point node 9400 (i.e. specific to our FE model) of LCL, medial-lateral position of the tibial component, anterior-posterior position of the femoral component, stiffness of the MCL ligament, x coordinate of the insertion node 9403 of MCL.

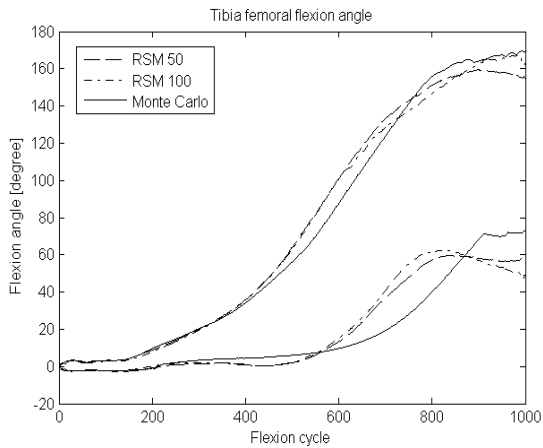
a) Tibio-femoral flexion angle (degrees)

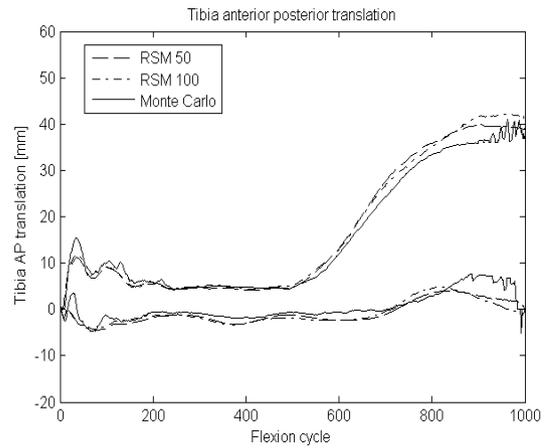
b) Tibial anterior-posterior translation (mm)

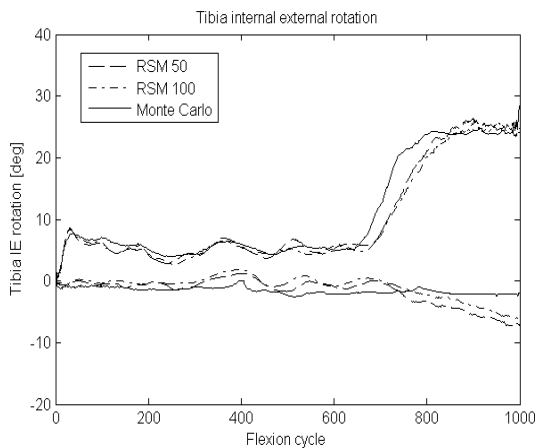
c) Tibial internal-external rotation (degrees)

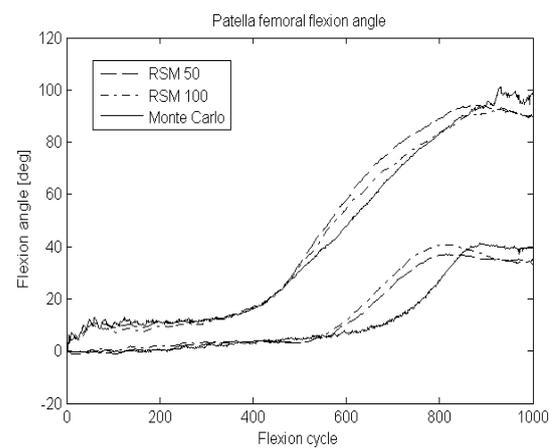
d) Patello-femoral flexion angle (degrees)

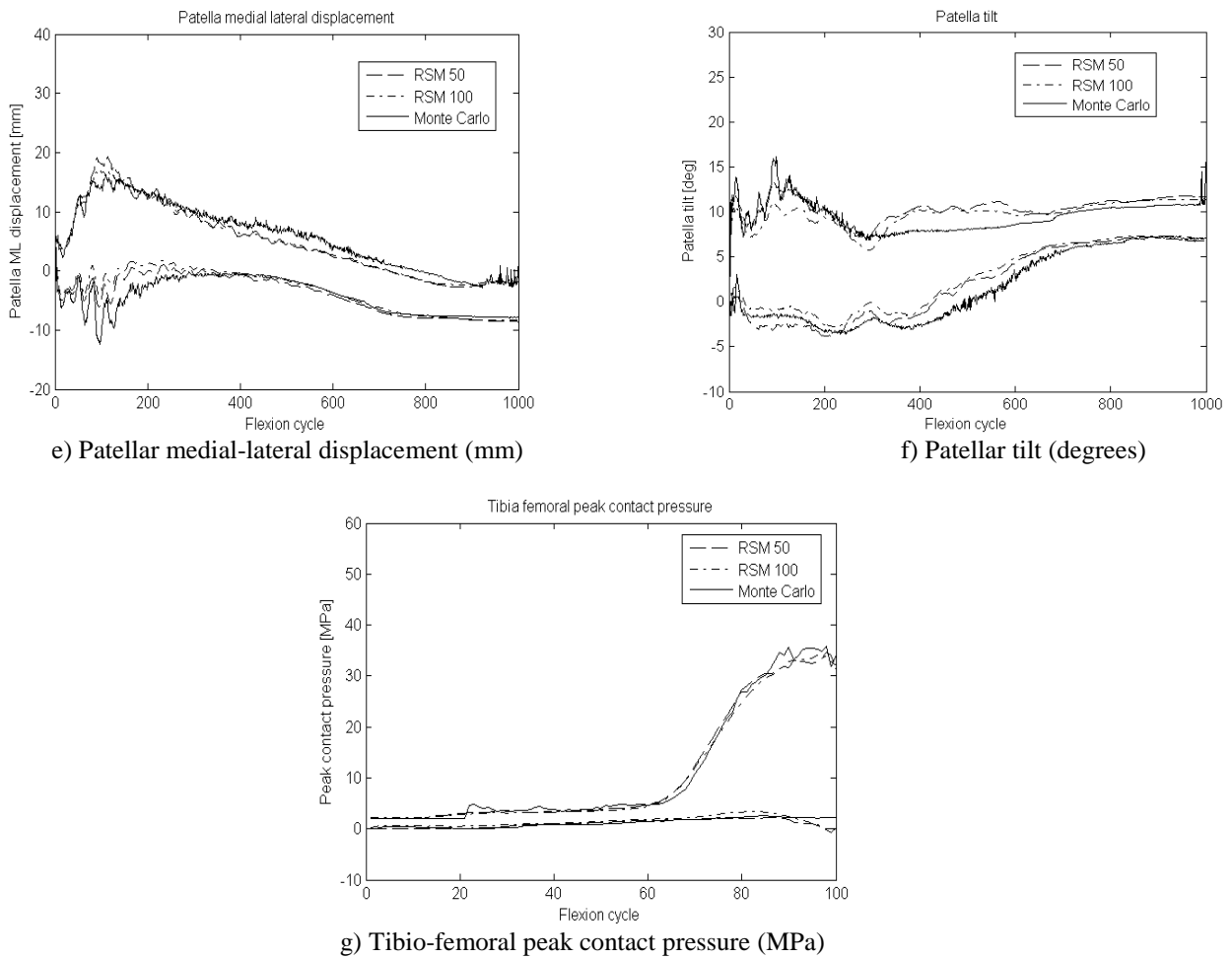

e) Patellar medial-lateral displacement (mm)

f) Patellar tilt (degrees)

g) Tibio-femoral peak contact pressure (MPa)

**Fig. 4.** Comparison between RSM model-predicted envelopes with 50 FE simulation points (dashdot line), 100 FE simulation points (dashed line), and MCST (solid line) with 800 points for the passive flexion FE model with the reduced set of 19 input variables.

It is important to underline for this paper that a comparison of this list with the similar list of 22 input key parameters obtained for the active flexion FE model [1,2] reveals that the quadriceps load, the quadriceps initial rotations and the friction of the leg-ankle and the pelvis-hip joints play a more important role in the passive flexion than in the active flexion motion. These results confirm similar findings from the literature regarding the relationship between the quadriceps tendon and the passive flexion movement of the human lower leg [28, 29].

Finally, two probabilistic FE studies were implemented with the reduced set of 19 input variables: a MCST with 800 FE simulation points (i.e. mean/percentile values became constant at 400 FE points) was run and the RSM with 50 and 100 FE simulations were used to calculate the envelopes (Fig.4). The envelopes of performance obtained with the 800 MCST FE simulations and the reduced set of 19 input variables compared very well with the envelope of performance obtained with the 800 MCST FE simulations with the full set of 78 variables and the envelopes of performance obtained with the RSM-50 and RSM-100 for the passive flexion FE model with the reduced set of 19 input variables. For example, the differences between the MCST with the full set of 78 input variables and 800 FE simulation points and the MCST with the reduced set of 19 input variables and 800 FE simulation points were 1.07 degrees for tibio-femoral flexion angle, 0.81 mm for tibial anterior-posterior translation, 2.44 degrees for tibial internal-external rotation, 2.46 degrees for patella-femoral flexion angle, 0.89 mm for patella medial-lateral displacement and 1.48 degrees for patellar tilt.

The differences between the MCST with the reduced set of 19 input variables and the RSM-50 with the reduced set of 19 input variables were of 5.82 degrees for tibio-femoral flexion angle, 1.24 mm for tibial anterior-posterior translation, 1.06 degrees for tibial internal-external rotation, 3.59 degrees for patella-femoral flexion angle, 1.17 mm for patella medial-lateral displacement, 1.11 degrees for patellar tilt, 1.6 MPa for tibio-femoral peak contact pressure. The differences between the MCST with the reduced set of 19 input variables and the

RSM-100 with the reduced set of 19 input variables were of 6.34 degrees for tibio-femoral flexion angle, 1.05 mm for tibial anterior-posterior translation, 1.44 degrees for tibial internal-external rotation, 3.81 degrees for patella-femoral flexion angle, 1.39 mm for patella medial-lateral displacement, 1.16 degrees for patellar tilt, 1.55 MPa for tibio-femoral peak contact pressure.

This suggests that a smaller number of FE simulations can be used in order to obtain the envelopes of performance based on the RSM-50/100 method and the reduced set of 19 input variables. However, the probabilistic FE analyses based on the MCST (i.e. with either the full set or the reduced set of input variables) represents the gold standard that it should be employed in every probabilistic FE analysis and in addition other probabilistic methods can be employed for exploratory studies.

In conclusion there were realized four probabilistic FE analyses: two probabilistic FE analyses based on MCST and RSM and based on the full set of 78 input variables and another two probabilistic FE analyses based on MCST and RSM and based on the reduced set of 19 input variables.

Further work will involve the development of a real-life decision support system for orthopaedics which to include the present results and the ones shown in [1,2] which were based also on a similar distributed computing paradigm.

ACKNOWLEDGMENT

The author would like to thank to European Union Framework Programme 6 (FP6) for financing this work under the project DESSOS (Decision Support Software for Orthopedics). The author would like to thank also to the following people for their support in this work Prof Mark Taylor, Dr Elena Samsonova, Dr. Magdalena Janosova, Dr Lucy Knight, Mrs Muriel Beaugonin, Mr. Pierre Guyon, Dr. Michael Strickland and Prof David Barrett.

REFERENCES

[1] C.T.C. Arsene and B. Gabrys, *Probabilistic finite element predictions of the human lower limb model in total knee replacement*, Medical Engineering and Physics, 35, 1116-1132, 2013.

[2] CT.C. Arsene, *Probabilistic finite element prediction of the active lower limb model*, Advanced Materials Research, vols. 463-464, 1285-1290, 2012.

[3] R. Huiskes, E.Y.S., Chao, *A survey of finite element analysis in orthopaedics biomechanics: the first decade*, J. Biomech., 1983, 385-409, 1983.

[4] P.J. Prendergast, *Finite element models in tissue mechanics and orthopaedic implant design*, Clin. Biomech., 12, No. 6, 343-366, 1997.

[5] M. Taylor, P.J. Prendergast, *Four decades of finite element analysis of orthopaedic devices: Where are we now and what are the opportunities*, Journal of Biomechanics, vol. 48(5), 767-778, 2015.

[6] P.S. Walker, A. Garg, *Range of motion in total knee arthroplasty*, Clinical Orthopaedics and Related Research 262, 227–235, 1991.

[7] L.D.Dorr, R.A. Boiardo, *Technical considerations in total knee arthroplasty*, Clinical Orthopaedics and Related Research 205, 5–11, 1986.

[8] C.K. Fitzpatrick, M.A. Baldwin, C.W. Clary, A.Wright, P.J. Laz, P.J.Rullkoetter, *Identifying alignment parameters affecting implanted patellofemoral mechanics*, Journal of Orthopaedic Research, 30(7):1167-1175, 2012.

[9] C.K. Fitzpatrick, M.A. Baldwin, C.W. Clary, A. Wright, PJ. Laz, PJ Rullkoetter, *Characterizing alignment parameters affecting patelofemoral TKR mechanics*, Journal of Bone & Joint Surgery, 94, 156-156, 2012.

[10] C.K. Fitzpatrick, C.W. Clary, PJ Laz, PJ Rullkoetter, *Relative contributions of design, alignment, and loading variability in knee replacement mechanics*, Journal of Orthopaedic Research, 30(12), 2015-2024, 2012.

[11] P.J. Laz, S. Pal, J.P. Halloran, A.J. Petrella, P.J. Rullkoetter, *Probabilistic finite element prediction of knee wear simulator mechanics*, Journal of Biomechanics, 39, 2303-2310, 2006.

[12] P.J. Laz, S. Pal, A. Fields, A.J. Petrella, P.J. Rullkoetter, *Effects of knee simulator loading and alignment variability on predicted implant mechanics: a probabilistic study*, J. Ortho.Res., 24, 2212-2221, 2006.

[13] D.P. Nicolella, B.H. Thacker, H. Katoozian, D.T. Davy, *Probabilistic risk analysis of a cemented hip implant*, ASME Bioengineering Division 50, 427-428, 2001.

[14] M.Browne, R.S. Langley, P.J. Gregson, *Reliability theory for load bearing biomedical implants*, Biomaterials 20, 1285-1292, 1999.

[15] F.H. Dar, J.R. Meakin, R.M. Aspden, *Statistical methods in finite element analysis*. Journal of Biomechanics 35(9), 1155-1161, 2002.

[16] J. Shi, M. Browne, M. Strickland, G.Flivik, M.Taylor, *Sensitivity analysis of a cemented hip stem to implant position and cement mantle thickness*. Comp. Meth. in Biom. and Biomed. Eng.17:1671-1684, 2014.

[17] M.E. Melis, E.V. Zaretsky, R. August, *Probabilistic analysis of aircraft gas turbine disk life and reliability*, Journal of Propulsion and Power 15(5), 658-666, 1999.

[18] R.E. Kurth, K.S. Woods, *Probabilistic damage tolerant analysis for fatigue critical aircraft components*, ASME, Aerospace Division 28, 89-97, 1992.

[19] Y. Zhang, Q. Liu, *Reliability-based design of automobile components*, Proceedings of the Institute of Mechanical Engineers, Part D. Journal of Automobile Engineering 216(6), 455-471, 2002.

[20] P. Jacob, L. Goulding, *An explicit finite element primer*, Glasgow, NAFEMS Ltd, 2002.

[21] A.C. Godest, M. Beaugonin, E. Haug, M. Taylor, P.J. Gregson, *Simulation of a knee joint replacement during gait cycle using explicit finite element analysis*, Journal of Biomechanics 35, 267-275, 2002.

[22] S.S. Isukapalli, S. Balakrishnan, P.G. Georgopoulos, *Computationally efficient uncertainty propagation and reduction using the stochastic response surface method*, 43rd IEEE conf. Decision and Control, Bahamas, US. 2004.

[23] https://www.cygwin.com/

[24] Strickland, MA, Arsene, CTC, Pal, S., Laz, PJ and Taylor M., *A Multi-Platform Comparison of Efficient Probabilistic Methods in the Prediction of Total Knee Replacement Mechanics*, Journal of Computer Methods in Biomechanics and Biomedical Engineering, Taylor and Francis, 2010, 1476-8259, Volume 13, Issue 6, 2010, pp. 701 – 709.

[25] Arsene, C.T.C., Strickland, M.A., Laz, P.J., Taylor, M., *Comparison of two probabilistic methods for finite element analysis of total knee replacement*, Proceedings of the 8th International Symposium on Computer Methods in Biomechanics and Biomedical Engineering, CMBBE 2008, February 2008, ISBN 978-0-9562121-0-8.

[26] PAM-OPT users manual. 75761 Paris Cedex, 16, France: ESI Group, Rue Hamelin, BP 2008-16, 2007.

[27] PAM-CRASH users manual, 75761 Paris Cedex, 16, France: ESI Group, Rue Hamelin, BP 2008-16, 2007.

[28] Jolles, BM, Garofalo, R., Gillain, L., Schizas, C., *A new clinical test in diagnosis quadriceps tendon rupture*, Ann R Coll Surg Engl, 89(3), 259-261, 2007.

[29] Hak, D.J., Sanchez,A., P., Trobish, *Quadriceps Tendon Injuries*, Orthopedics, vol. 33(1), ISSN:0147-7447, 2010.